*Essay*

# Nearly Forgotten Cosmological Concept of E. B. Gliner

**Dmitry Yakovlev * and Alexander Kaminker**

Ioffe Institute, Politekhnicheskaya 26, 194021 St. Petersburg, Russia; kam@astro.ioffe.ru
* Correspondence: yak@astro.ioffe.ru

**Abstract:** E. B. Gliner started his scientific career in 1963 at the age of 40. In 1965, when the existence of the cosmological constant $\lambda$ seemed unnecessary to most cosmologists, he renewed interest in the problem by emphasizing a material interpretation of de Sitter space (i.e., the space curved in the presence of $\lambda$). According to that interpretation, the curvature is produced by a cosmological vacuum (now identified as dark energy of the universe). In 1970, Gliner proposed a description of exponential expansion (or contraction) of the universe at the early (or late) evolution stage dominated by cosmological vacuum. In 1975, Gliner (with I. G. Dyminikova) suggested a model of the early universe free of Big Bang singularity, and developed a scenario of nonsingular Friedmann cosmology. Many of these findings were used in the modern inflation scenarios of the universe, first proposed by A. A. Starobinsky (1979) and A. Guth (1981) and greatly multiplied later. However, these inflation scenarios differ from the scenario of Gliner and Dymnikova, and Gliner's contribution to cosmology is nearly forgotten. The history and the essence of this contribution are outlined, as well the difference from the inflation theories.

---

## 1. Introduction

This paper is dedicated to the memory of Erast Borisovich Gliner (26.01.1923–16.11.2021), a remarkable person and scientist whose life history was, unfortunately, very dramatic. He started his scientific career in 1963, at the age of 40, at the Ioffe Institute in Leningrad (now St. Petersburg). The earliest stages were the most successful, with his proposal to treat the initial state of universe by de Sitter vacuum world [1]; a similar cosmological vacuum is now considered as a valid model for dark energy. In addition, he individually [2], and together with his student I. G. Dymnikova [3], built a model for nonsingular Friedmann cosmology that differs from the inflation scenarios of the early universe suggested later (starting from [4,5]). However, his working conditions at the Ioffe Institute became difficult, and he emigrated to the USA in 1980. Unfortunately, he was unable to find a place where he could fully focus on cosmology, and his contribution to cosmology is currently almost forgotten.

Here, we attempt to remind the reader about this contribution. Section 2 outlines his life before the start of his scientific work; this period is important for understanding his personality. In Section 3, we give a brief outline of Gliner's work at the Ioffe Institute; this period was most successful from a scientific point of view, but not so pleasant. Section 4 summarizes the American period of his life. Finally, we conclude in Section 5.

We (DY and AK) were colleagues (but certainly not the friends) of E. Gliner, working in the same Department of Theoretical Astrophysics of the Ioffe Institute from the end of the 1960s until Gliner's emigration to the USA. He was a very gentle, ironic, insightful and extraordinarily talented individual. He had a gift of explaining the most difficult scientific problems in simple terms.

There were rumors that his life had been terrible during World War II and afterwards, but he was not eager to discuss it with us. We actually learned his life story about one decade after his departure to the USA.

E. Gliner had two unique features. First, he was friendly and open to anyone; he was conflict-free, ready to establish and develop useful working relations. Second, he was rock solid in what he considered to be matters of principle. This characteristic brought him a lot of harm.

## 2. Before the Ioffe Institute (Before 1963)

### *2.1. Before the War*

E. Gliner was born in Kiev on 26 January 1923. All his official documents show the wrong birth date, 3 February 1923, because of the delayed registration of his birth in violation of existing rules; the recorded birthday was assumed to satisfy the rules. Erast's father, Boris Gliner, left the family after he learned of the expected child, and Erast never met him. His mother, Bella Rubinshteyn (1889–1989), was a bacteriologist. Erast was her only child.

In 1926, the family moved to Leningrad (now St. Petersburg). In the summer of 1940, Erast graduated from school (Figure 1a) and was admitted to the Chemistry Department of Leningrad State University (LSU). He was deeply interested in chemistry. His second dream was to study the theory of General Relativity (GR).

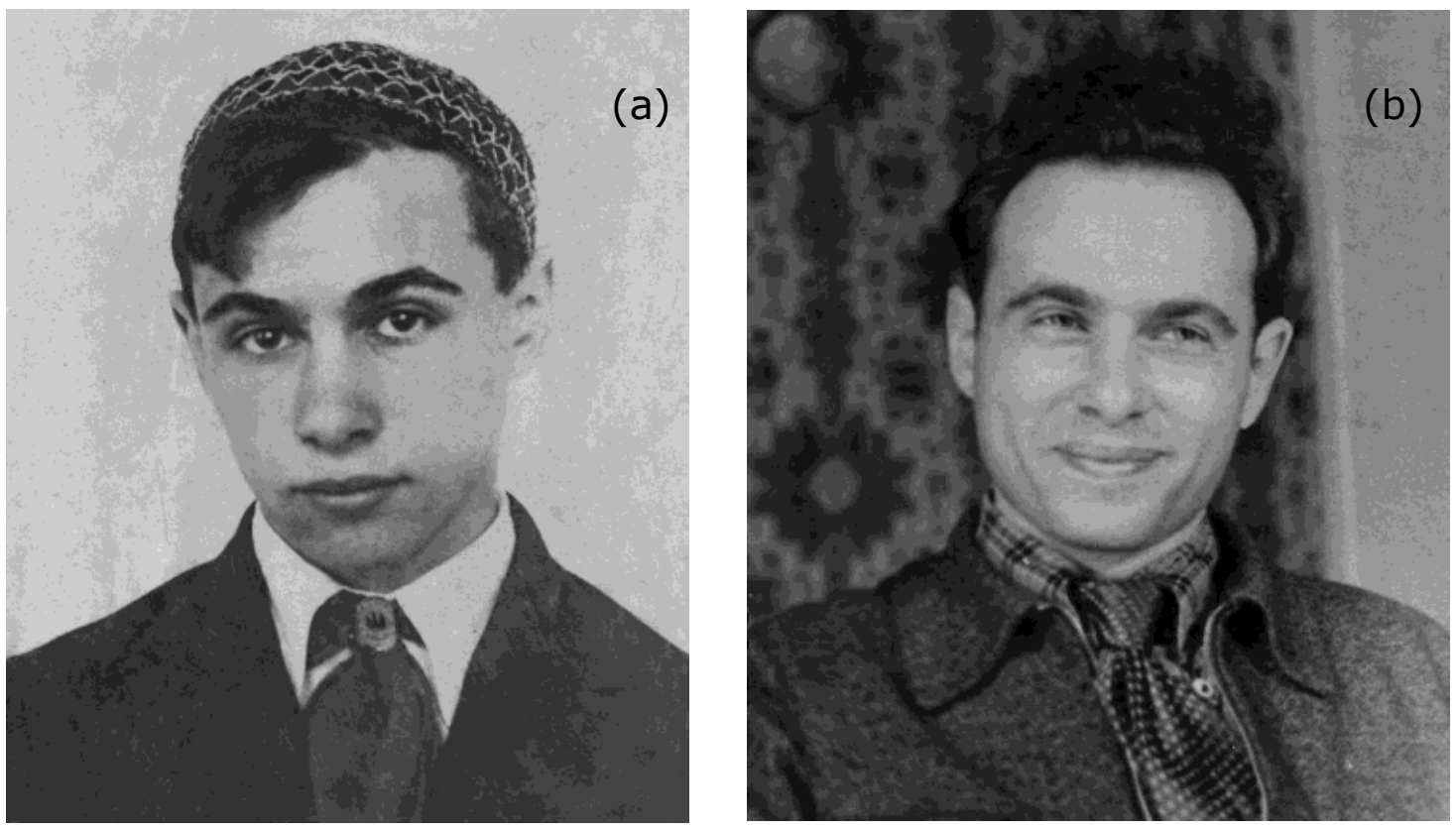

**Figure 1.** (**a**) 1940. Farewell, school years. (**b**) 15 June 1954. Released after imprisonment, but full of energy and optimism. In Tolmachevo, more than 100 km from Leningrad, where he was allowed to reside. From family archive; courtesy of Arkady and Bella Gliner.

### *2.2. The War and Perception of Beauty*

On 22 June 1941, the Second World War came to Russia. Erast finished his first year at LSU and was mobilized to build fortifications near Leningrad. Students worked hard to build a military aerodrome and almost finished, but in September, the Germans started a powerful attack, and the students were ordered to return to Leningrad. They ran, but the Germans were faster. Erast was shell-shocked on his way to Leningrad, then carried and thrown into a waiting train. By some miracle, the train delivered him to Leningrad. He slowly recovered and spent the coldest and hungriest winter of 1941–1942 in Leningrad. He suffered from the terrible starvation that was typical of Leningrad citizens during that winter (with about 100,000 deaths from hunger each month). In April of 1942 he was evacuated to Saratov.

His medical condition prevented him from enlisting, but the situation at the front was terrible. At the end of April, for patriotic reasons, he insisted on enlisting and was assigned to artillery. His fellow soldiers liked him. He fought in Ukraine and Belarus as a common soldier and as a sergeant. He was awarded the Order of the Red Star. He was wounded three times. The last wound to the right arm on 30 October 1943 required amputation and he was hospitalized for several months. He was then released from the army and returned to Leningrad in the summer of 1944.

For medical reasons, he could not work with chemicals and was unable to continue his chemistry education. He was admitted to the Physics Department of LSU. He had to start from the first year, because the programs of the Chemistry and Physics Departments were different.

Additionally in 1944, Erast married Galina Ilchenko (a chemist). In 1945, their daughter Bella was born. To make ends meet, he took a part-time job in December 1944 as a physics teacher at an evening school for working people.

After the nightmares of the war, studying at LSU was easy; classes were not always attended. The half-forgotten beauty of Leningrad fascinated Erast and made him think about the aesthetic perception of beauty (architectural monuments, literature, painting, music). It seemed (and indeed it is so!) that aesthetic perception does not begin with a political assessment, but with an artistic quality. Erast expressed his ideas in the House of Writers at the literary circle hosted by B. D. Chetverikov, quite a known writer. Several circle members were present.

In those times, such ideas were dangerous. Six people, including E. Gliner, were arrested on 13 March 1945; they barely knew each other. The case was conducted by the Department of the Ministry of State Security in the Leningrad District. The war was still going on, so the case was judged by a military tribunal under the laws of war. The arrested people were interrogated at night using torture. They were charged with "participation in the activities of an anti-Soviet group" with the aim of destroying Soviet and Communist Party literature. On 19 May, E. Gliner was sentenced by the military tribunal under Articles 58-10, part 2 and 58-11 to 10 years with subsequent five-year suppression of civic rights and no right to appeal.

Gliner was meant to be sent to a special prison as disabled. Until then, he was kept in the prison named Kresty and situated in the center of Leningrad. The captain of the Ministry of Internal Affairs, S. F. Pivovarov, took pity on him, and recategorized him from a prisoner–disabled to a prisoner–specialist (because of one completed university year). Such specialists worked in military construction bureaus. Although the conditions there were close to an ordinary prison, these bureaus were considered a little bit "merrier".

Until the summer of 1952, Gliner worked at the Special Construction Bureau (OKB-172) in Kresty designing naval gun systems. He participated in several projects and showed outstanding mathematical ability. For one of the developments in OKB-172, he obtained a copyright certificate dated 1953. In 1952, he was transferred to Moscow, to KB-1, not far from the Sokol metro station. One of the two chief engineers of KB-1 was the young and rapidly advancing Sergo (Sergey) Beria, the son of the almighty Lavrenty Beria. The bureau developed various types of communications and anti-aircraft missile systems. E. Gliner showed himself as a talented designer and proposed a "radical technical improvement" of the control system (receiving a certificate dated 1954).

Routine life of KB-1 was interrupted in March 1953 by the death of J. Stalin. In June 1953, L. Beria was arrested; the prisoners of KB-1 were dispersed. E. Gliner was sent to a camp near Krasnoyarsk (the coldest place in Russia). At first, it was bad there: a slender man of Jewish origin with an amputated right arm met various prisoners, not only political ones but also criminals and traitors (e.g., police officers under the German occupation of Russia). However, everything worked out reasonably well. Gliner was eventually transferred to the Special Technical Bureau (OTB-1) of Yeniseystroy, where he designed heating systems for buildings in the far north.

There were even certain advantages in Krasnoyarsk: due to the extremely cold climate, a credit system was used that reduced the actual sentence term. Thanks to this, Gliner was released on 25 April 1954. He returned to Leningrad but was legally forbidden to live there for five years. He lived, therefore, in Tolmachevo, more than 100 km from Leningrad (Figure 1b). Soon after, fate finally smiled: unexpectedly, the Military Collegium of the Supreme Court canceled the verdict as a result of newly discovered evidence (an unusual case in those times). It is unclear how this was discovered, but it was determined that B. D. Chetverikov (the organizer of that literary circle) was a secret informer of state security. In addition, he was also the main witness for the prosecution at Gliner's trial, which was prohibited by law. As a result, the verdict was thrown out and a retrial was ordered. Finally, on 5 August 1955, E. Gliner was fully exonerated due to lack of evidence.

*2.3. P.O.691 and LSU*

With one year of university education, his past as a political prisoner and his Jewish origin, finding a job was difficult. However, even before the full exoneration, in May 1955, Gliner was offered a permanent job at a construction bureau (Leningrad branch of KB-1), and he accepted the position. After the War, these bureaus had such unbearably bleak names that they were commonly known by the number of the local post office, in this case P.O.691. These bureaus were allowed to hire former prisoners with unexpunged conviction.

E. Gliner was already over 30, and his prospects were gloomy. He had no place of his own to live. In 1957, a second child was born in the family, his son Arkady. The opportunity to conduct real scientific research was postponed until the summer of 1963. Despite all the problems, life was quiet and satisfying. Gliner's main efforts were focused on work and university education.

E. Gliner worked at P.O.691 until February 1961. He was respected and quickly moved up, ending as chief of the theoretical (computational) department. Since he was conflict-free and easily communicated with many people, he was often sent on business trips to Moscow requesting funds and equipment. His trips were successful. His salary was decent.

In October 1955, E. Gliner was readmitted to the Physics Department of LSU. Unfortunately, the documents confirming that he had completed the first academic year in 1944–1945 were lost. He was forced to start again from the first year (for the third time!). It was difficult to combine study and his P.O. job. He could not attend university classes. He learned course material by himself, and then passed exams as an external student. However, his preparations were thorough, and grades were always excellent. Twice, in 1958 and 1959, he had to take leaves of absence from LSU for urgent P.O. jobs. From 1961, by a special agreement with V. A. Fock, he was allowed to be a full-time student on a free schedule.

Two additional events should be mentioned. The first was associated with L. D. Landau, the famous Soviet theoretical physicist. As already mentioned, working as a prisoner in KB-1, E. Gliner demonstrated extraordinary scientific potential. One freelance mathematician employed at the same KB-1 advised Gliner to contact Landau after his release. The mathematician promised to recommend Gliner, and he kept his word. After the release, during his business trips to Moscow, E. Gliner attended several Landau seminars, and talked to Landau after one seminar. Landau arranged a meeting at which he briefly evaluated Gliner. It is well known that Landau was usually harsh and sarcastic to students. However, he was extremely polite to Gliner, "almost gentle" (probably because Gliner's arrest reminded him of his own arrest on 29 April 1938 and a year spent in prison). Landau explained that to join his team, Gliner should pass the theoretical minimum exams, for which Gliner was apparently not yet ready. But—continued Landau—he was ready to make an exception for Gliner and take him on the condition that he would pass the first exam within one year. Unfortunately, this proposal was not acceptable to Gliner for many practical reasons, and he politely refused.

The second significant event took place in 1962. In Moscow, the Publishing House of "Vysshaya Shkola" published a textbook "Differential Equations of Mathematical Physics" [6] co-authored by N. S. Koshlyakov, E. B. Gliner and M. M. Smirnov. Gliner was the main co-author. A serious motive for writing was a need for money to improve his living conditions. N. S. Koshlyakov (1891–1958) was an outstanding mathematician, arrested in Leningrad on a false accusation at the end of 1941. Gliner met him in prison and greatly respected him. It was a matter of principle to make him the first author. The textbook was a great success and was published four times in Russia, and translated into English and Japanese.

In 1961 Gliner's P.O.691 was reorganized. This allowed him to take a long leave of absence to complete his university studies. Officially, he was expected to return to the P.O., but this did not happen. By the middle of 1963, he had passed all the exams and defended his thesis "Investigation of the singularity of Schwarzschild's external solution." In June 1963 he graduated from LSU (diploma with honors). B. P. Konstantinov (1910–1969), then director of the Ioffe Institute, somehow heard about E. Gliner. He sent a request to LSU with the proposal to employ him. In this way, Gliner was admitted to the Ioffe Institute.

## 3. At the Ioffe Institute (1963–1981)

### *3.1. What to Do: Math or GR?*

That was the end of E. Gliner's pre-Ioffe-Institute life. He was already over 40, when the most active creative period of scientists usually ends. With Gliner, it was just the start. He only managed, through enormous efforts, to obtain a university education. He probably thought that his main life difficulties were over, and he could fully focus on scientific research. However, unfortunately, it was only the start of new problems.

On 1 August 1963, E. Gliner was officially admitted as a trainee researcher with a monthly salary of RUB 100 (about five times less than what he earned at P.O.691). Then, he needed to choose the direction of his research. For practical reasons, it would be better to defend his PhD as soon as possible. The simplest solution would be to choose mathematical physics. He was already a co-author of a successful textbook on mathematical physics; there was a strong Department of Mathematical Physics at the Ioffe Institute, where Gliner would be most welcome. However, for him, it was matter of principle to choose GR, his second dream since school (after chemistry). E. Gliner's first two years at the Ioffe Institute were the most fruitful.

### *3.2. On the Cosmological Constant*

In modern language, the main interests of E. Gliner focused on studying the cosmological vacuum and the evolution of the early universe. That was closely related to Einstein's cosmological constant $\lambda$ and the vacuum de Sitter space. To summarize the contribution of E. Gliner to this science, let us outline the development of ideas on the cosmological constant.

A. Einstein proposed his first version of field equations in 1915. However, he was dissatisfied with them because they could not describe a static universe (that was assumed natural by most scientists in those times). Accordingly, in 1917 Einstein [7] modified his equations by including the so-called $\lambda$ (cosmological) term to be able to have a static universe as a solution. His final equations read

$$R_{ik} - \frac{1}{2} R\, g_{ik} + \lambda g_{ik} = \kappa T_{ik}, \tag{1}$$

where $i, k = 0, 1, 2, 3$ enumerate coordinates (0 for a time-like coordinate; 1, 2, 3 for space-like ones); $g_{ik}$ is the metric tensor (with squared differential proper distance interval $ds^2 = g_{ik}\, dx^i\, dx^k$), $R_{ik}$ is the Ricci tensor, $R$ is the scalar curvature, $T_{ik}$ the energy–momentum tensor, and $\kappa = 8\pi G/c^4$ is the Einstein gravitational constant (with $G$ being the ordinary gravitational constant and $c$ the speed of light). In the simplest case of hydrodynamical motion of perfect fluid, one has $T_{ik} = (P + \epsilon)u_i u_k - P g_{ik}$, where $P$ is the pressure, $\epsilon$ is the energy density of the fluid, and $u_i$ denote its 4-velocity components.

With application of (1) to cosmology, one needs $P$ and $\epsilon$ of "normal" matter (ordinary + dark). The cosmological term (the last term on the left-hand side) is determined by $\lambda$. In the Einstein theory [7], $\lambda$ is a fundamental constant. Presently, one often considers solutions to (1) in which $\lambda$ becomes a dynamical variable; it can be nearly constant in a restricted space–time region but may vary outside it.

Additionally, in 1917, de Sitter [8] presented a thorough analysis of (1) including the $\lambda$-term. He took the tensor $\lambda g_{ik}$ on the left-hand side (which, according to Einstein, describes geometrical properties of space–time), formally introduced the tensor $T^{\text{v}}_{ik} = \lambda g_{ik}/\kappa$ and moved it to the right-hand side (that describes material properties of space–time). This transforms (1) into

$$R_{ik} - \frac{1}{2} R\, g_{ik} = \kappa \tilde{T}_{ik} \equiv \kappa (T_{ik} + T^{\text{v}}_{ik}). \tag{2}$$

It looks similar to the Einstein equations without the $\lambda$-term but with the renormalized energy–momentum tensor $\tilde{T}_{ik}$. With this trick, geometrical properties of the $\lambda$-term trans-form into material properties. Accordingly, the presence of $\lambda$ is equivalent to the appearance of some extra (vacuum) energy density $\epsilon_v$ (or corresponding effective mass density $\rho_v = \epsilon_v/c^2$) and vacuum pressure $P_v$, with

$$\epsilon_v = \rho_v c^2 = -P_v = \lambda/\kappa. \quad (3)$$

Then, the total (renormalized) pressure $\tilde{P}$ and energy density $\epsilon$ become

$$\tilde{P} = P + P_v, \quad \epsilon = \epsilon + \epsilon_v. \quad (4)$$

Moreover, de Sitter [8] considered a special world in which $P = \epsilon = 0$ [$T_{ik} = 0$ in (1)]. No "normal" matter is present there but space–time is curved (and the curvature is constant, being determined by $\lambda$, the only one parameter of the theory). Such a world is well known as the de Sitter world. For the most interesting case of $\lambda > 0$, one has positive $\rho_v$ but negative $P_v$. The de Sitter world is infinite in size, uniform, isotropic, (quasi) static, Lorentz-invariant; all reference frames are equivalent there. The negative (contracting) pressure corresponds to negative *inertial* mass density $\rho_{veff} = \rho_v + 3P_v = -2\rho_v$. It produces an effective repulsion (antigravitation) that works against attractive Newtonian gravitation and ensures the (quasi)stationarity of space.

It is well known that the metric of the de Sitter space can be written in the static form

$$ds^2 = \left(1 - \frac{r^2}{r_v^2}\right) c^2\, d\tau^2 - \left(1 - \frac{r^2}{r_v^2}\right)^{-1} dr^2 + r^2(d\vartheta^2 + \sin^2\vartheta\, d\varphi^2), \quad r_v = \sqrt{\frac{3}{\kappa\epsilon_v}}. \quad (5)$$

Here, $\tau$ is a time-like coordinate, $r$ is circumferential radius, $\vartheta$ and $\varphi$ are ordinary spherical angles, and $r_v$ is an event horizon of the causally accessible sphere of the de Sitter's space ($r < r_v$). This metric covers such a sphere, with the center in any point of de Sitter world. It is singular at the horizon $r = r_v$ but the singularity is removable. One can describe the same world using other coordinates, but the metric will become non-stationary and cover different regions of the world.

The history of the cosmological constant was similar to a thriller (see, e.g., Petrosian [9], who described the period before the early 1970s). In the 1920s, there appeared growing evidence that our universe was not static but evolving. The necessity of a static universe became illusive. In 1922 and 1924, Friedmann [10,11] laid the foundation for his (Friedmann's) cosmology at an arbitrarily fixed $\lambda$. However, soon after, it became clear that the universe definitely demonstrates Hubble expansion at such a rate that observations were easily explained assuming $\lambda = 0$.

During the long period from the 1930s to the end of the 1990s, there were no reliable data on the presence of the cosmological constant. It was a long era of "stagnation" when $\lambda$ was not needed for interpreting observations. Even A. Einstein considered his $\lambda$-term as a mistake (an unnecessary complication of GR). Many excellent monographs on GR and cosmology published in those times (e.g., Synge [12], Weinberg [13], Misner et al. [14], Zeldovich and Novikov [15]) described the Einstein field Equation (1) with the $\lambda$ term and the de Sitter world, but then stated that the $\lambda$ term was not needed and dropped it in most chapters.

### 3.3. Heavy Vacuum

We believe that the interest in the cosmological vacuum was renewed by E. Gliner in 1965, in his first paper [1]. It was published in Russian in ZhETF (Zhurnal Eksperimentalnoi i Teororeticheskoi Fiziki). The English translation (in JETP—Journal of Experimental and Theoretical Physics) appeared in 1966. ZhETF was the best Soviet journal in those times, and it was difficult for an unknown author to be published there. It seems to be the most important of Gliner's publications, a basis for his dissertation and subsequent research. In modern language, the paper was devoted to the state of the very early expanding universe (or late collapsing universe), which was assumed to be the de Sitter world.

Gliner [1] reviewed the de Sitter space that had been traditionally studied, postulating some geometrical curvature ($\lambda$ term). In contrast, Gliner considered it as the space filled by the cosmological vacuum that produced the curvature. That vacuum was not just an empty space, but the space filled by a special substance with the energy density $\epsilon_v$, mass density $\rho_v$ and the negative pressure $P_v$, in accordance with Equation (3). Gliner called its state "vacuum-like" and its material medium heavy vacuum. The state was specified by the parameter $\epsilon_v$. The heavy vacuum was spread uniformly over the space. In the case of some admixtures of "normal" matter, the vacuum could interact with them only via gravitational forces.

Specifically, Gliner [1] used the main properties of the de Sitter world (Lorentz invariance, equivalence of all inertial references frames) and basic principles of GR. In his own elegant manner, he rederived the first expression in Equation (3), $\epsilon_v \equiv P_v$. The second expression just trivially relates $\epsilon_v$ and $\lambda$. Let us stress that according to Petrosian [9],

Equation (3) was known and discussed among a narrow circle of GR experts long ago, particularly, in the 1930s; the equivalence of geometrical and material approaches was also debated. Gliner recognized the equivalence but strongly preferred material interpretation.

Let us note that Gliner's paper [1] was refereed by Ya. B. Zel'dovich, as acknowledged in the text. Zel'dovich did not approve Gliner's approach in those times (and possibly considered it as too simple). However, he did not reject the article but sent it for revision and then let it go.

Looking back at those times from now, it is easy to understand why the first paper [1] by an unknown author was left unnoticed outside the former Soviet Union. However, it was noticed by the elite of Soviet science. There were eminent scientists who highly approved (including V. A. Fock, A. D. Sakharov, V. L. Ginzburg) and disapproved (including Ya. B. Zel'dovich, V. N. Gribov). The paper was resonant in the Soviet Union and attracted attention to the problem of cosmic vacuum. Since Zel'dovich was a reviewer and had not published anything about the cosmic vacuum before, it would be natural to assume that his attention was somewhat attracted as well.

In 1967, two years after the publication of [1], the interest in the cosmological constant was renewed, triggered by observations of distant quasars (see, e.g., [15] for more details). Initially, some observational data looked as if they could be explained by assuming nonzero $\lambda$ but that turned out to be a false alarm. Although it was false, it lasted enough time to attract the attention of cosmologists. In addition, once attracted, the attention was growing: there was no way back to the $\lambda = 0$ times, first for theorists, and then for observers as well.

Following observations of distant quasars, Zel'dovich in 1967 suggested a theoretical explanation of these observations assuming nonzero $\lambda$. In his brief communication [16], he discussed a relationship of cosmological constant with the physics of elementary particles. In 1968, Zel'dovich [17] published a review paper on "Cosmological Constant and Elementary Particles." It triggered the second wave of interest in the cosmological problem. It was not local—many scientists over the world were involved; it has been discussed later in numerous review papers and textbooks. Indeed, its aims were most ambitious and irresistible–to unite studies of elementary particles and their interactions with cosmology.

Let us add that the physical nature of cosmological vacuum remains unclear even now (in 2022). The role of the cosmological constant in the very early universe is a subject of debate. Modern cosmology is rich in different and controversial models that cannot be fully tested by current observations.

Nevertheless, many observations unambiguously show the presence of the cosmological vacuum (dark energy) at the present Friedmann expansion stage. Convincing evidence was provided by observations of cosmologically distant type Ia supernovae, which resulted in the award of the Nobel prize for physics to S. Perlmutter, B. Schmidt and A. Rice in 2011 (see a press release of the Royal Swedish Academy of Sciences https://www.nobelprize.org/uploads/2018/06/press-14.pdf, accessed on 4 October 2011). In the present epoch, the energy density of dark energy constitutes 69% of the total energy density (with the rest distributed between the dark matter, about 27%, and the ordinary matter). There is growing evidence that this dark energy is distributed uniformly, being nearly constant in space and time (e.g., Abbott et al. [18], and references therein), just as the heavy vacuum advocated by Gliner [1]. According to observations and theoretical interpretation with the Λ CDM model, its density is $\rho_v \approx 6 \times 10^{-30}$ g cm$^{-3}$, and its pressure $P_v \approx -\epsilon_v$; the corresponding cosmological constant is $\lambda \approx 10^{-56}$ cm$^{-2}$. Its main observational manifestation is to accelerate the universe expansion due to antigravitation. At early Friedmann expansion stages, the heavy vacuum density $\rho_v$ constituted smaller fraction of the total mass density; the vacuum acceleration was slower than the deceleration of the universe expansion due to Newtonian attraction of gravitating matter. Accordingly, the overall expansion decelerated. The deceleration stopped and was followed by acceleration about 5.4 billion years ago. Now the acceleration prevails, and its importance will increase in the future. One usually treats the present-day dark energy as a leftover of the initial dark energy, meaning that $\rho_v$ was time-dependent and much larger in the very early universe.

### *3.4. Scaling Factor in a Vacuum-Dominated World*

There is another early Gliner publication [2] that deserves special attention. It was published in 1970 in Reports of the USSR Academy of Sciences (communicated by A. D. Sakharov). To discuss it, let us remind the reader about the Friedmann cosmology of the expanding universe, which is based on the Robertson–Walker metric (e.g., [15]) for the uniform and isotropic universe,

$$ds^2 = dt^2 - \frac{dr^2}{1 - \mathrm{k}\,(r/a)^2} - r^2\,(d\vartheta^2 + \sin^2\vartheta\, d\varphi^2). \qquad (6)$$

Here, $t$ is cosmological time and $r$ is circumferential radius. All physical quantities, such as pressure and density, depend only on $t$. Index k = 0 corresponds to a spatially flat universe, while k = 1 or −1 refers to a closed or open universe. The universe evolution is determined by a scale factor $a = a(t)$. From the Einstein Equation (2), one obtains two well-known Friedmann equations for $a(t)$,

$$\ddot{a} = -\frac{\kappa c^2}{6}\left(\tilde{\epsilon} + 3\tilde{P}\right) a, \quad \dot{a}^2 = \frac{\kappa c^2}{3}\tilde{\epsilon} a^2 - c^2\mathrm{k}. \qquad (7)$$

Dots mean time derivatives; $\tilde{\epsilon}$ and $\tilde{P}$ are given by (4).

For example, let us take a standard flat (k = 0) Friedmann model without $\lambda$ term ($\tilde{\epsilon} = \epsilon$, $\tilde{P} = P$) at the radiation-dominated expansion stage ($P = \epsilon/3$). Equations (7) immediately give a well-known Friedmann solution $a(t) \propto \sqrt{t}$. This means that we can take some epoch ($t = t_0$) and choose a matter element at a distance $r_0$. According to (6), $r = r_0$ labels the same element at other times, but the physical radius of the element changes as $a(t) = r_0 \sqrt{t/t_0}$.

Now, we return to Gliner's paper [2]. Actually, Gliner mostly focused on the collapse of a massive body stopped by the appearance of heavy vacuum, but he also discussed the inverse problem. To make it more logical, we reformulate his results for the case of the de Sitter space with a heavy vacuum as an initial stage of universe. A more detailed cosmological scenario was described in the next publication [3] but he temporarily focused on the very early stage dominated by the cosmological vacuum. The de Sitter world is known to be quasi-static, being actually unstable to small perturbations. This should produce further expansion accelerated by antigravitation and accompanied by the creation of "normal" matter.

The initial amount of "normal" matter was assumed to be small. Gliner studied its evolution following the motion of test particles in a de Sitter environment. To this aim, he used metric (6) and Equation (7) for the scale factor $a(t)$ assuming $\tilde{\epsilon} = \epsilon_v$ and $\tilde{P} = -\epsilon_v$. It is easy to show that in this case (at any k = 0, ±1) $a(t)$ satisfies the equation $\ddot{a} - (c/r_v)^2 a = 0$, where $r_v$ is the de Sitter world horizon given by (5). Accordingly, $a(t)$ is a superposition of two exponents; one is growing with $t$ while the other decaying. Clearly, it is sufficient to leave the first term for a qualitative analysis. Assuming $a(0) \sim r_v$, one obtains

$$a(t) \sim r_v \exp(tc/r_v). \tag{8}$$

Please note that the scale factor $a(t)$ refers to a reference system of "normal" matter (even in the case of very small energy density of this matter). Equation (8) means exponentially rapid expansion of newly born "normal" matter in the vacuum-dominated world. Please note that such rapid expansion does not violate GR criteria of subluminality (e.g., [19]). Let us add that E. Gliner doubted the correctness of using the above reference frame at the vacuum-dominated stage throughout his life (e.g., Gliner [20]); we do not share his doubts.

Presently, the exponential expansion (8) is basic for modern inflation theories. The priority of Gliner in deriving this expression was acknowledged, for instance, by Linde [21]. Inflation theories started to appear (approximately) in the decade after Gliner's publication [2].

### *3.5. Dissertation*

The dissertation of E. Gliner was completed in 1970. It was called "On Extremely Dense State of Physical Medium" and was based on five articles. The first article [1] has already been discussed above. Two others [22,23] were devoted to possible generalizations of Einstein field equations in terms of the rank 4 generalized energy–momentum tensor. This subject will be developed further as outlined in Section 4. The next paper [24] was a study of removable singularities in GR, and the last paper [2] was also described above. In addition, there were summaries of reports at two conferences. E. Gliner was the only author of all publications. There is no doubt that all results of the dissertation were obtained solely by E. Gliner.

Before going on, let us note that obtaining a scientific degree in Russia is a two-step process. The first step is to become a Candidate of Sciences (similar to PhD) and the second step (often called *doctoral*) aims to become a Doctor of Sciences (similar to Habilitation in some countries). A defense at the first step required two reviewers (opponents), with three opponents at the next step. In any case, one needed to present the dissertation text and its summary, as well as some other official documents. Defenses were conducted at Special Defense Counsels in research institutes or universities; thesis summaries were posted over a net of similar counsels in advance.

In the first half of 1971, preparation for PhD defense was going well. Two opponents were found—the world-famous scientists and academicians A. D. Sakharov and V. A. Fock. Gliner had known Fock from his university years. He had first learned of Sakharov by reading his scientific publications. Later, they met accidentally during one of Gliner's business trips to Moscow, and then were meeting occasionally. Gliner would always say that he was only developing Sakharov's ideas (e.g. [25]). Sakharov believed that Gliner was one of the best experts in GR. Both opponents were sure that the dissertation could be defended as doctoral, skipping the PhD stage.

Regular pre-defense exams were successfully passed in May and early summer of 1971. Nothing presaged troubles, and where would they come from? Those times in the former Soviet Union were reasonably quiet and favorable for Gliner. He was a disabled and decorated veteran of the World War II, repressed and rehabilitated. He was fully protected by Soviet laws unless he was involved in actions against the political system. However, with his bitter experience in the past, he could not even imagine this.

### *3.6. Troubles*

He could not imagine this, but it happened. The problem was with his assumed opponent A. D. Sakharov. The political disfavor of Sakharov for upholding democratic principles in the former Soviet Union had already started. Sakharov's review of Gliner's thesis was "reclassified" into a political affair and became dangerous for Gliner.

The disaster came in the second half of 1971. What was done was fully illegal and realized by illegal methods; it was kept secret. According ly, it is next to impossible to reconstruct exact dates, but the sequence of events is clear. Two main participants from the Ioffe Institute administration were director, V. M. Tuchkevich, and scientific deputy director, N. V. Fedorenko. There is no doubt that the illegal (oral) order came from much higher circles. Scientific secretary of Ioffe Institute, G. V. Skornyakov, who was in charge of PhD documents, was inevitably involved as well. Several colleagues close to Gliner, primarily L. E. Gurevich, were also familiar with the case, but that is all.

Gliner dealt with N. Fedorenko, who insisted on two points. First, Sakharov is removed from the list of opponents, in which case the defense is allowed, perhaps even as doctoral; otherwise, all dissertation documents are taken away by administration and there will be no defense at all. Second, to frighten Gliner, it was said that the GR theory (Gliner's topic) was not included in the work plan of the Ioffe Institute; accordingly, the topic should be changed, otherwise Gliner could lose his job.

It was unbelievable cruelty, but nobody protested because almost nobody knew. As for E. Gliner, it was a matter of principle not to disregard Sakharov. He refused, and his dissertation documents were locked away at the institute.

The threat of job loss seemed disastrous. E. Gliner decided to seek protection from famous scientists. He called Sakharov, but he was not in Moscow (and how could he help?). Then he called V. L. Ginzburg, whom he did not previously know. Ginzburg listened and helped by including GR in work plans for scientific studies of the USSR Academy of Sciences. Moreover, V. L. Ginzburg supported Gliner's work for many years ahead.

Further events were dramatic. By the end of December 1971, N. Fedorenko became seriously ill. He died on 1 February 1972.

### *3.7. Defense of the Forbidden PhD*

Then, according to personal notes of E. Gliner (courtesy of Arkady and Bella Gliner), the idea appeared to organize a defense disregarding the Ioffe Institute administration. A. D. Sakharov contacted influential Estonian scientist Harald Keres and told him the story (however, according to [38], it was L. E. Gurevich who contacted Keres). Keres was interested and invited Gliner to give a seminar talk at Tartu University.

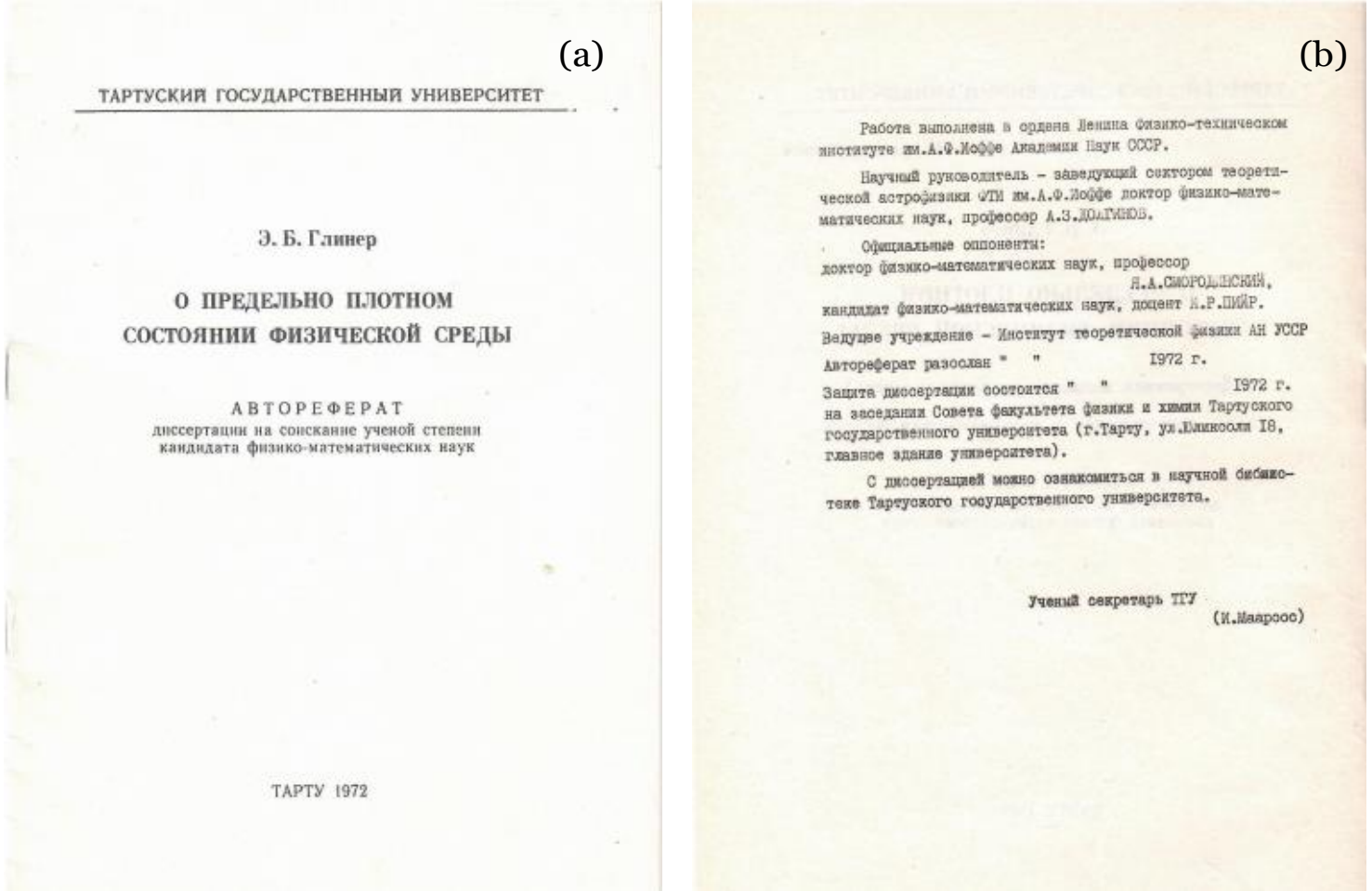
(a)

ТАРТУСКИЙ ГОСУДАРСТВЕННЫЙ УНИВЕРСИТЕТ

Э. Б. Глинер

О ПРЕДЕЛЬНО ПЛОТНОМ СОСТОЯНИИ ФИЗИЧЕСКОЙ СРЕДЫ

АВТОРЕФЕРАТ
диссертации на соискание ученой степени кандидата физико-математических наук

ТАРТУ 1972

(b)

Работа выполнена в ордена Ленина Физико-техническом институте им.А.Ф.Иоффе Академии Наук СССР.

Научный руководитель – заведующий сектором теоретической астрофизики ФТИ им.А.Ф.Иоффе доктор физико-математических наук, профессор А.З.ДОЛГИНОВ.

Официальные оппоненты:
доктор физико-математических наук, профессор Я.А.СМОРОДИНСКИЙ,
кандидат физико-математических наук, доцент И.Р.ПИЙР.

Ведущее учреждение – Институт теоретической физики АН УССР

Автореферат разослан " " 1972 г.

Защита диссертации состоится " " 1972 г. на заседании Совета факультета физики и химии Тартуского государственного университета (г.Тарту, ул.Юликооли 18, главное здание университета).

С диссертацией можно ознакомиться в научной библиотеке Тартуского государственного университета.

Ученый секретарь ТГУ (И.Маароос)

**Figure 2.** Cover page of Gliner's dissertation summary (in Russian). (**a**) The author's name, PhD title, defense location (Tartu University) and year: 1972. (**b**) Names of scientific adviser, and both reviewers. See the text for details. From archive of DY.

The talk was successful, and Keres suggested Gliner defend in Tartu. The only problem was to retrieve the dissertation documents from the Ioffe Institute. This retrieval was made by G. V. Skornyakov, scientific secretary. Once the defense was arranged and the thesis summary was published, the administration had no legal ways to interfere: the ban was illegal.

As a result, a successful defense took place on 2 June 1972 at a meeting of the Council of Physics and Chemistry Department at Tartu University (Figure 2a,b). A defense of the doctoral dissertation was not possible there, because the Council in Tartu could consider only PhD theses. Moreover, the Higher Attestation Commission forbade Sakharov from being an opponent. The official opponents were Ya. A. Smorodinsky (employed at the Joint Institute for Nuclear Research in Dubna) and I. R. Piir from the Tartu University. The (formal) supervisor was A. Z. Dolginov, the head of the Theoretical Astrophysics Department of the Ioffe Institute. Incorrect statements that Sakharov was the reviewer can still be found in the literature and on the Internet.

After the defense, the Ioffe Institute administration had no rights to punish Gliner for his "disobedience," but it was easy to make his stay at the Institute uncomfortable. This finally contributed to the emigration of E. Gliner to the USA in 1980.

### *3.8. Nonsingular Friedmann Cosmology*

After the defense, E. Gliner (Figure 3) continued developing his ideas, although it was not so pleasant. Nevertheless, he obtained new important results.

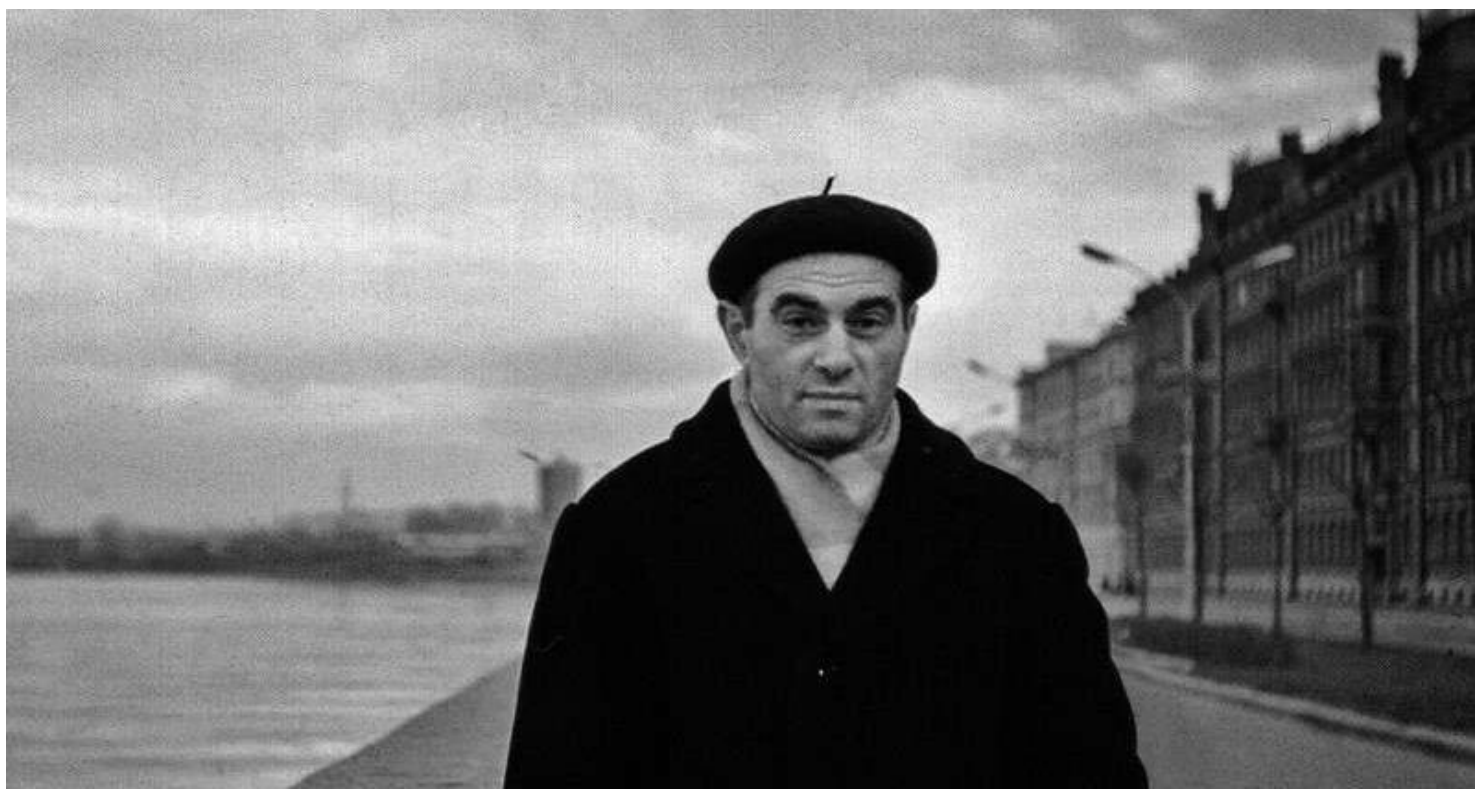

**Figure 3.** Early 1970s. Embankment of the Neva river in Leningrad (now St. Petersburg). From Gliner family archive; courtesy of Arkady and Bella Gliner.

The most striking was a model of generalized Friedmann cosmology [3] published in 1975. It was made in collaboration with I. G. Dymnikova, Gliner's only student. The idea was to describe evolution from the initial stage of the universe to the later Friedmann expansion stage using a simple uniform and isotropic model. All unknowns, such as energy density and pressure, were functions of cosmological time $t$ (see Section 3.4).

Gliner and Dymnikova viewed the initial universe as a de Sitter world containing a heavy vacuum ([1], see also Sections 3.2 and 3.3). Such a world is quasi-static; its energy density $\epsilon_{vi}$ ("i" labels the initial stage because later $\epsilon_v$ depended on $t$) is the largest; $\epsilon_{vi}$ was taken as a free parameter where nature was not specified. By construction, the model contains no singularity (infinite energy density) in the initial state, an attractive advantage over standard singular Big Bang models. A solution of nonsingular cosmology [3] could be, in principle, continued into the past (beyond the initial stage). It could have been, for instance, a previous universe, where its collapse had been stopped by heavy vacuum repulsion and followed by subsequent expansion.

In [3], as in [2], the expansion was thought to be triggered by fluctuations of a generically unstable heavy vacuum. Gliner and Dymnikova focused on the evolution of one universe, which was initially confined in a sphere of horizon radius $r_v$ in the de Sitter world. Their consideration was limited by a causally connected region. Since the entire world was infinite, neighboring spheres could form other universes from the same initial de Sitter world.

The authors considered a toy model with two stages in one selected universe. The first (vacuum-dominated) stage started at $t = 0$ and lasted as long as $\tilde{\epsilon}(t) \geq \epsilon_1$, where $\epsilon_1$ is a parameter of the model. At this stage, in accordance with (4), $\tilde{\epsilon}(t) = \epsilon_v(t) + \epsilon(t)$ and $\tilde{P}(t) = P_v(t) + P(t)$, with $\epsilon_v(0) = \epsilon_{vi}$ and $P_v(t) = -\epsilon_v(t)$. It was assumed that $\epsilon_v(t)$ decreased with time and became negligible by the end of the first stage, $t = t_1$. The second stage was approximated by the Friedmann expansion stage where the vacuum contribution was neglected.

At the first stage, the authors assumed a simple phenomenological relation

$$\tilde{P} = -\tilde{\epsilon} + \frac{4}{3}\epsilon_1[(\epsilon_{vi} - \epsilon)/(\epsilon_{vi} - \epsilon_1)]^{\alpha}, \tag{9}$$

parametrized by a factor $\alpha$, with $0 \leq \alpha \leq 1$. It describes an increase of $P_v$ from $-\epsilon_{vi}$ at $\epsilon_v = \epsilon_{vi}$ to $P_v = 0$ at $\tilde{\epsilon} = \epsilon_1$, with $P_1 = \epsilon_1/3$; heavy vacuum decays creating "normal" matter. In the limit of $\alpha \to 0$ it corresponds to a step function with $P_v = -\epsilon_{vi}$ at $\epsilon > \epsilon_1$ and $P_v = 0$ at $\epsilon = \epsilon_1$ (meaning first-order phase transition at $t = t_1$). In any case, at $t = t_1$ the authors imposed a transition from the vacuum-dominated expansion to the standard Friedmann expansion.

One can question the study of the vacuum-dominated epoch. The restriction of consideration to causally connected region was postulated; other possible assumptions were not discussed. Because of initial heavy vacuum fluctuations, spherical symmetry can be broken and the initial expansion can be a random (rather than regular) process. Moreover, the equation of state $\tilde{P}(\epsilon)$ was regarded as given and temperature-independent.

By construction, the Gliner–Dymnikova solution cannot reproduce extremely large inflation of the universe size predicted by current inflation theories. The latter inflation is realized via outflow far beyond the horizon; but these distances are unavailable in [3] although the driving Equation (8) is the same. An expansion in [3] is rather insensitive to a specific equation of state (to a specific value of parameter $\alpha$).

In summary, Gliner and Dymnikova [3] seemed to build the first cosmological model with variable vacuum density $\rho_v(t)$ and avoided Big Bang singularity. Removal of the singularity produced a burst of emotions and scatter of opinions at a seminar in Moscow that preceded publication of the paper [3]. The paper was extraordinarily short: two pages with two references in the list. It is cited in the literature, but rarely, and mostly "by repetition." It is so compressed that current readers can hardly comprehend its essence. Let us stress that the model was clearly formulated and correctly solved using the formalism similar to that in inflation theories (but imposing horizon restrictions). It is a pity that its status has been largely misunderstood; note once more that inflation models were unavailable in 1975.

More detailed and clear description of the same model was given by Dymnikova [26] in 1986 when inflation theories were blooming. If Gliner and Dymnikova [3] used the results of [26] for $\alpha = 0.5$ with $a_1 \sim 10^2$ cm and $t_1 \sim 2 \times 10^{-32}$ s (which corresponded to $\rho_{vi} \sim 2 \times 10^{73}$ g cm$^{-3}$ and $\rho_1 \sim 10^{71}$ g cm$^{-3}$), they would obtain $r_v \sim 10^{-23}$ cm and $\rho_1 \sim 10^{71}$ g cm$^{-3}$. By $t = t_1$ their universe would strongly expand in size by a factor of $a_1/r_v \sim 10^{25}$ (although $\rho$ drops only by a factor $\sim 200$). Such an expansion is huge according to ordinary standards. Nevertheless, it is "too slow" for current inflation theories which predict enormous factors $\sim 10^{800}$ due to expansion beyond horizon (see, e.g., Linde [21]).

We should add that in 1975 L. Gurevich [27] also suggested a model of the early universe born from a heavy vacuum. He based it on the same results [1,2] and considered somewhat different aspects of the problem.

### *3.9. Gravitational-Wave Experiment*

In 1977, E. Gliner was noticeably cheered up by promises to conduct a gravitational-wave experiment at the Ioffe Institute. He was offered to be PI, while his colleague, G. M. Gorodinsky, agreed to supervise technical implementation. From the very start, the offer looked suspicious, as if someone made it with the intention to destroy everything later. The project lasted several months at the stage of preliminary theoretical study. Based on this activity, a 74-page report on "Gravitational-Wave Experiment with a Broadband Gravitational Antenna" was written in 1978; moreover, a new type of monochromatic gravitational-wave detector was proposed [28]. Then the project was closed.

### 3.10. *Seminar on Non-Organic Chemistry*

In 1978–1979, rumors were spread that the Institute administration had finally decided to promote E. Gliner to the position of senior researcher, which meant recognition of scientific merits and a noticeable increase in salary. For a long time, Gliner was junior researcher (there were no other positions except juniors and seniors), which did not correspond to his scientific potential and was completely unfair. The possibility of becoming senior would greatly improve if he defended his doctoral dissertation. However, the sad story of his PhD defense gave him little chance.

Nevertheless, it was announced that his case would be an exclusion. He was offered an opportunity to give a talk at a seminar of the Theoretical Department on the condition that the talk would be on a topic that Gliner had not previously dealt with. If the talk was successful, the promotion was promised.

The seminar took place and (in the opinion of DY, who was present) it was exhilarating. The subject was non-organic chemistry. Gliner talked as if he had studied complex non-organic molecules all his life, easily coped with cumbersome chemical terminology, and tremendously enjoyed the subject himself. Obviously, his childhood passion for chemistry had not been forgotten. However, no promotion followed.

### 3.11. *Emigration*

There were other "little jokes" of the same type which finally led to Gliner's decision to emigrate. In addition, his children wanted this, believing that in the West they could improve their lives.

The only opportunity for E. Gliner to leave the country was to emigrate to Israel. As a result, in the fall of 1979 he requested from the Ioffe Institute a certificate to start the procedure, and on 20 November he obtained it. On 2 April 1980, he resigned from the Ioffe Institute although he actually had not attended it since autumn 1979. He left the Soviet Union on 18 May 1980 with his mother, daughter, granddaughter and son, officially, for Israel (flying through Vienna). Instead, the family stayed in Italy until they flew to San Francisco on 18 August 1980. Galina, Gliner's wife, could not leave the country at the same time. To allow the others to emigrate, the couple divorced in August 1979. She came to the USA and joined the family a year later.

### 3.12. *Inflation*

When E. Gliner was retiring from the Ioffe Institute and settling in the USA, a great breakthrough occurred in cosmology—inflationary models of expanding universe were invented. The first models were proposed by A. A. Starobinsky [4] (from the USSR) and A. Guth [5] (from the USA).

These models use a metric close to the de Sitter metric and demonstrate a very strong expansion ("inflation") of the universe before the transition to the Friedmann expansion stage. These key elements are similar to those suggested earlier by Gliner but there were differences discussed in Section 3.4 and mentioned below.

Very soon, the inflationary cosmology became extremely popular; it is now accepted by most cosmologists. It includes theories of a different nature that are sometimes diverse. The field is developing; old versions are updating. There is even no strict consensus which theories should belong to this class.

In any case, inflation cosmology is a new and important instrument based on the contemporary theory of fields and their interactions. Additionally, it uses the entire arsenal of modern physics of elementary particles. As a rule, inflationary theories of the early universe assume the presence of a scalar field $\phi$ that mediates the structure and dynamics of the universe. Its nature is still (2022) not known. For instance, one mentions hypothetical inflatons. It can be coupled to other fields which enlarges inflation scenarios.

A theory usually deals with a scalar field $\phi$ and "normal" matter in the early universe; both constituents may vary in space and time. It is assumed, as usual, that the "normal" matter consists of ultra-relativistic particles with huge thermal energies $\sim k_B T$; its pressure $P = \epsilon/3$, where $k_B$ is the Boltzmann constant, $T$ is the temperature, and the energy density $\epsilon \propto T^4$.

The field $\phi$ is treated as a dynamical variable which evolves according to field theory (within some $\phi$ range) being regulated by some effective potential $V(\phi)$; this potential can depend also on $T$. The presence of temperature means thermalization, which can be not full, allowing quasistationary states, e.g., supercooled liquid. For a given local element, the field $\phi$ can take different values, but the most probable value corresponds to a minimum of $V(\phi)$. In addition to the absolute minimum corresponding to full thermalization, there could be other local minima corresponding to quasistationary states with higher $V(\phi)$. Depending on prehistory, a system may appear not in the absolute minimum, but in a quasistationary state.

The universe evolution can be followed even from Planck times $\sim 10^{-43}$ s. Inflation theories predict the appearance of special stages at which $\ll V(\phi)$, where $\phi$ is a quasistationary local minimum. This state appears to be similar to a vacuum-dominated state that is approximately described by the de Sitter world. For instance, it could be a supercooled state where the temperature is relatively low; this case was considered by Guth [5] in 1981. Then, the universe suffers exponential inflation (8) and the transition to thermodynamically stable state. In many scenarios, the inflation stage occurs at $t \sim 10^{-35}$–$10^{-32}$ s. At this stage, $\epsilon_v$ stays almost constant; nearly exponential inflation carries the expanding matter beyond the event horizon and decreases temperature (that will be increased later).

Evidently, the vacuum-dominated inflation era in these theories has common features with the vacuum-dominated stage of the cosmological scenario of Gliner and Dymnikova [3]. In both cases, there is a large expansion of the universe in size, accompanied by a weak variation of the energy density $\epsilon$. Nevertheless, there is a basic disagreement explained in Section 3.8: the expansion in [3] is restricted by a horizon and appears not too strong, while the inflation theories allow for much stronger expansion beyond horizon.

Inflation theories (e.g., [21,29,30]) have the potential to resolve several fundamental problems of contemporary cosmology, such as Big Bang singularity, flatness of the universe, the horizon problem, baryon asymmetry, non-detection of primordial Dirac monopoles, etc.

Let us stress once more a great scatter of various inflation theories. Their common feature is a vacuum-dominated stage with exponential expansion. For instance, the idea of Starobinsky [4] in 1979 was to arrange that stage over Planck times, $10^{-42}$ s, in the very early (close to Big Bang) universe by including quantum corrections to Einstein equations. The effective mass density of that heavy vacuum would be $\sim 10^{93}$ g cm$^{-3}$, and the horizon size would be $\sim 10^{-33}$ cm.

## 4. After the Ioffe Institute (1981–2021)

### *4.1. America*

E. Gliner liked America and felt completely free there. Eventually, the family bought a house in San Francisco on Faxon Avenue.

He enjoyed driving. He and Galina drove over almost all of America, visiting many picturesque places. It was convenient in the car: as a disabled person and World War II veteran, he could park everywhere.

In 1983, his paper [31] (in collaboration with I.G. Dymnikova) was published in Physical Review D. It was actually written in Leningrad, but was not published in the Soviet Union due to a chain of unlucky circumstances, and the original was lost when Gliner left the country. He reconstructed the text by memory. The paper proposed a covariant description of energy in GR. The approach was based on introducing a generalized en-

ergy–momentum tensor of the 4th rank. After the publication, Ya. B. Zeldovich called Gliner and said that he liked the article and agreed with it. He confessed also that he had underestimated the scientific potential of E. Gliner and his colleague.

However, it was difficult for an immigrant scientist in his 60s to find a permanent job in the United States. E. Gliner managed to find temporary jobs several times (Joint Institute for Laboratory Astrophysics, University of Colorado, Boulder; McDonnell Center for Space Science, Washington University, St. Louis). To his regret, the work was mainly related not to GR but, for example, to solar physics or calculating the trajectories of celestial bodies. He became the author and co-author of more than a dozen publications on the physics of the Sun and the solar corona. Studies in GR were also encouraged, but as supplemental ones. As far as we know, Gliner was mostly surrounded by friendly and knowledgeable scientists. Among them, we would like to mention V. A. Osherovich, who tried to support him for many years. However, the lack of communication with colleagues directly involved in GR and cosmology was disappointing.

He did not lose optimism, and engaged himself in journalism and human rights activities. When A. D. Sakharov was in exile in Gorky (now Nizhny Novgorod), E. Gliner wrote articles in his defense. Two of them were published in Nature [32,33].

*4.2. Unemployment, Institute for Theoretical Studies and Seminar at Stanford*

His last temporary job ended in 1995, when Gliner fell seriously ill. Gradually, his health improved (mainly through self-care because decent medical insurance was unavailable). He was unable to find another job.

He tried to work at home, which was not easy because his access to scientific libraries and electronic resources was limited (to almost nothing). Nevertheless, he was working and submitting articles (e.g., [20]). He often used his home address and added the affiliation "Institute for Theoretical Studies," or "ITS" for short. Articles were published in that way, and reports appeared on the Internet that he was employed at this institute.

E. Gliner's last attempt to make contacts with American cosmologists was made in 1997. A. Linde, a famous cosmologist (originally from the N. P. Lebedev Institute in Moscow), kindly invited him to give a seminar talk for a very influential group of cosmologists at Stanford University (not far from San Francisco), and Gliner gladly accepted.

From the very beginning of inflationary cosmology, Gliner considered himself as its participant (at the early pre-inflation stage). However, internally, he questioned exponentially strong inflation of universe size obtained using a reference frame comoving with "normal matter" at the vacuum-dominated stage [see Section 3.4; Equation (8) and the discussion around]. His doubts were summarized later in [20].

Accordingly, he devoted his talk at Stanford to criticism of inflationary models. Naturally, it was not a great success. Being in the USA, Gliner was isolated from the GR-community which prevented him from fully comprehending achievements of inflation cosmology. As for the audience, they were convinced of the largely recognized inflation theories. As a result, there appeared a wall of misunderstanding which was never broken.

*4.3. The (Almost) Final Verdict*

Unfortunately, the contribution of E. Gliner to cosmology is gradually becoming forgotten by the most active part of the astrophysical community. He started too early, in 1965, when the cosmological constant was widely thought to be unavailable. He investigated the effects of the heavy vacuum in the very early universe before the appearance of the first inflation theories.

Undoubtedly, his results have been used for developing current cosmological concepts, but the field is growing so fast that those old involvements have been buried under new achievements.

As for Gliner, he was rather indifferent to publicity. Let us cite a few sentences from the recommendation letter of V. L. Gunzburg to E. Gliner (dated December 17, 2003; from the Gliner archive, courtesy of Arkady and Bella Gliner): "Gliner's personality combines a mild sense of humor with indifference to publicity. The promotion of his works has been really left to the mercy of fate. This poor practice has severely sacrificed Gliner's career but also let him concentrate on fundamental issues, i.e., has been reasonable to some extent."

*4.4. Late Recognition*

Being in the USA, E. Gliner was not fully isolated at all. He also felt some wave of support, produced mostly by his former compatriots.

The most amusing was a chain of honorary visits to his home in San Francisco. Many prominent physicists from the former Soviet Union and Russia considered it as their pleasant duty to visit him and stay there. Among them, we can mention A. D. Sakharov, V. L. Ginzburg, B. P. Zakharchenya (from the Ioffe Institute, a co-discoverer of exciton).

On January 23, 2003, a conference in his honor (his 80th birthday) was organized at the Ioffe Institute in St. Petersburg. It was his only visit to Russia after the emigration in 1980. There were many scientists, his good friends, many discussions and friendly conversations. His talk was most interesting and bright. Then, he visited Moscow for a few days and returned home.

As for scientific recognition, we have already described the late but most valuable support by Ya. B. Zel'dovich. Additionally, let us cite a piece of his review paper devoted to E. Gliner (written with another famous cosmologist, A. A. Starobinsky [34]). It clearly formulates the positive opinion of two world-renowned experts on Gliner's work. Here it is:

"As all this discussion [in 1960s] about the small nonzero $\lambda$ was continuing, the next step was made: Gliner [1] (1965) put forward the bold hypothesis (effectively contradicting the spirit of physics at that time) that the equation of state of matter $P = P(\epsilon)$ approaches the desired relation $P = -\ \epsilon$ at some very large energy density $\epsilon_0$ ..." And then Starobinsky and Zeld'ovich continued: "Later, Gliner [2] applied his hypothesis to the early state of the universe (1970) and naturally, obtained the de Sitter solution . . . at the initial state of the universe. This "hydrodynamical" inflation was further investigated in papers by Gliner and Dymnikova [3] (1975) and Gurevich [27] (1975) where several important observations were made including the possibility of multiple creation of individual Friedmann universes from one maternal de Sitter state."

We should also stress the indispensable support by V. L. Ginzburg during all active periods of Gliner's life. In a review article "What problems of physics and astrophysics seem especially important and interesting now (30 years later, already on the threshold of the 21st century)?", Ginzburg [35] emphasized the contribution of E. Gliner in advancing the problem of a cosmological vacuum in [1].

Somewhat later, Ginzburg offered Gliner the opportunity to write a review paper for UFN (Physics–Uspekhi). He asked him to describe the current state of cosmology of the early universe and compare various theories. Just then, new observations of distant type Ia supernovae appeared, confirming the effect of antigravity caused by the present-day cosmological vacuum.

Gliner gladly accepted, but had neither time nor energy to review modern inflationary cosmologies; they had grown up too widely. Instead, he wrote an essay [20]. It was provided with an editorial foreword by V. L. Ginzburg. At the end of the essay, as an appendix, a paper by Gliner and Dymnikova [3] on nonsingular Friedmann cosmology was reprinted.

That was the last published article by E. Gliner. Over six pages, he tried to compare his cosmological scenario with inflationary scenarios. He presented the same arguments as at the Stanford seminar in 1997. He stated that too rapid an expansion of an inflating universe is produced due to improper choices of reference frame at the vacuum-dominated stage. In contrast, we think that the main difference between Gliner–Dymnikova and inflation cosmological concepts comes from the Gliner–Dymnikova assumption that the expanding universe is limited by a casually connected region (see Sections 3.4, 3.8 and 3.12 for more details). Both concepts represent correct solutions of different problems (using the same basic elements but different assumptions).

We should also mention the strong support of E. Gliner by A. D. Chernin, who had known Gliner since Ioffe Institute times. He wrote several papers clarifying Gliner's results and their correspondence to observations (e.g., [36,37]). Gliner's results and biography are also described in a monograph by Silbergleit and Chernin [38].

It is a comfort to add that many problems discussed by E. Gliner have been further developed by I. G. Dymnikova (as reviewed, e.g., in Dymnikova [39]).

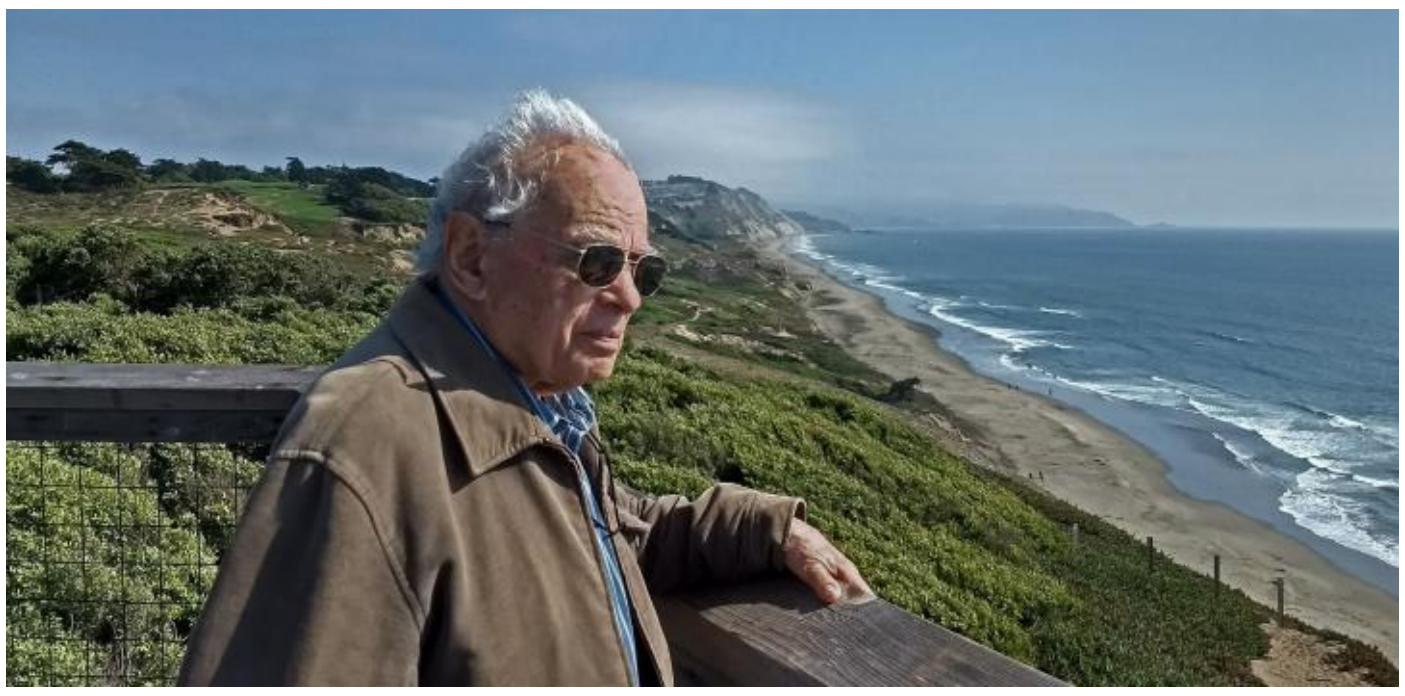

**Figure 4.** August 2017, San Francisco. Eternity of Pacific. From family archive; courtesy of Arkady and Bella Gliner.

### *4.5. Last Years*

The last years of Erast Gliner were non-eventful. On 28 August 2006, his wife Galina passed away; this was a great loss to him. He remained at the same house on Faxon Avenue. His children and grandchildren took care of him but without connections to other scientists he felt ageing and lonely (Figure 4). He died on 21 November 2021.

## 5. Conclusions

It is difficult to summarize the basic scientific achievements of Erast Gliner better than the summary by Starobinsky and Zeldovich [34] (Section 4.4).

Gliner's first paper [1] (1965) attracted the attention of cosmologists to the problem of cosmological (heavy) vacuum (Section 3.3). Gliner treated it as a special state of medium with huge energy (now called dark energy) and negative pressure that produces antigravitation. Gliner assumed that a heavy vacuum could be especially important in very early universe (at the beginning of rapid expansion) or in a late (collapsing) universe. That assumption was made when the cosmological constant was thought to be non-existent.

Later, Gliner [2] (1970) and Gliner and Dymnikova [3] (1975) (Sections 3.4 and 3.8) developed a scenario of the very early universe that initially contained a heavy vacuum; that vacuum was supposed to be generically unstable with respect to transformation to "normal" matter. Gliner [2] discussed the possibility of exponential expansion (8). Gliner and Dymnikova [3] detailed their cosmological model in which the initial universe contained no Big Bang singularity, the heavy vacuum decayed into "normal" matter, and the expansion could be accompanied by the production of many universes. All that work was the basis for modern cosmology and done before the appearance of inflation theories.

We have tried to show how hard Gliner's life was. Even his stay in Leningrad during the terrible siege winter (1941–1942), his participation in the war and 10-year imprisonment for anti-Soviet "perception of beauty" would be more than enough for one life (Section 2). However, he was always internally strong, full of energy, and demonstrated his extraordinary talents. He managed to graduate from university and was invited to join the Ioffe Institute (Section 3). All main scientific results were obtained there, but those years were not exactly pleasant, and he emigrated to the USA (Section 4). Although he found some temporary positions there, he was unable to continue his studies of GR and cosmology.

It is our aim to reflect on the legacy of this remarkable, gentle and friendly individual and extraordinarily talented scientist. It was not his fault that fate gave him too little time for active research. He accomplished it not thanks to but contrary to his fate. If he had been more fortunate, he could have done much more. It seems he recognized this, and did not spend his time on minor problems but focused (perhaps intuitively) on those which turned out to be fundamental. His contribution to cosmology is almost forgotten, which is unjust and unfair. It is a common assertion that "time puts everything in its place," and we wish it to become true.

**Author Contributions:** Both authors contributed equally. All authors have read and agreed to the published version of the manuscript.

**Funding:** This research was conducted in frame of Work Program 0040-2019-0025 of Ioffe Institute.

**Institutional Review Board Statement:** Not applicable

**Informed Consent Statement:** Not applicable

**Data Availability Statement:** The data underlying this article will be shared on reasonable request to the corresponding author.

**Acknowledgments:** We are indebted to those who provided us with most valuable information on E. Gliner, his life and his science during long period of data collection for this article. Our special thanks go to Bella, Arkady and Linda Gliner, D. A. Baiko, V. V. Ivanov, V. A. Osherovich, M. I. Patrov, E. Romm, N. N. Rosanov and V. A. Rubakov.

**Conflicts of Interest:** The authors declare no conflict of interest.

## Abbreviations

The following abbreviations are used in this manuscript:

| | |
|---|---|
| GR | General Relativity (theory) |
| LSU | Leningrad State University (now St. Petersburg State University) |
| UFN | Uspekhi Fizicheskikh Nauk (Russian original of Physics Uspekhi) |
| ZhTF | Zhurnal Experimentalnoi i Toeretichaskoi Fiziki (in Russian) |
| JETP | Journal of Experimental and Theoretical Physics (Engl. transl. of ZhETF) |
| ZhTF Pisma | Pisma v Zhurnal Experimentalnoi i Toeretichaskoi Fiziki (in Russian) |
| Pisma Astron. Zhurnal | Original Russian version of Soviet Astronomy Letters |